\documentclass{amsart}

\usepackage{graphicx}

\newtheorem{theorem}{Theorem}[section]

\theoremstyle{definition}
\newtheorem{definition}[theorem]{Definition}
\newtheorem{proposition}[theorem]{Proposition}

\newtheorem{example}[theorem]{Example}

\theoremstyle{remark}

\numberwithin{equation}{section}

\begin{document}

\title{Delayed Blockchain Protocols}

\author{Drew Stone}
\address{University of Pennsylvania, Mathematics Department}
\email{drstone@seas.upenn.edu}
\email{drew@thebairegroup.com}


\begin{abstract}
Given the parallels between game theory and consensus, it makes sense to intelligently design blockchain or DAG protocols with an incentive-compatible-first mentality. To that end, we propose a new blockchain or DAG protocol enhancement based on delayed rewards. We devise a new method for imposing slashing conditions on miner behavior, using their delayed rewards as stake in a \textit{Proof of Work} system. Using \textit{fraud proofs}, we can slash malicious miner behavior and reward long-lived, honest behavior.
\end{abstract}

\maketitle

\section{Introduction}
The hype around blockchains and DAGs (which we refer to interchangeably) has outlived their ability to compete against centralized service providers. The payment processors and money services businesses still scale more efficiently and more cheaply versus their blockchain competitors. This, however, has not stopped the rise in research around scalability and security-first designs. By building with new scalable and secure consensus protocols, these networks stand to realize the potential of decentralized cryptocurrencies. New designs, however, introduce new attack vectors. They change up the incentive structure of the underlying game by potentially reducing costs of attacks. With new scalable designs comes the hope for new game theoretic techniques for dealing with uncertain consequences.

The parallels between game theory and consensus, as exhibited by the repeated nature of mining for rewards, represent clear indicators in the usefulness of game theoretic techniques towards protocol design. Using inspiration from Proof of Stake protocols, especially slashing conditions, we develop a methodology for designing similar incentive mechanisms over Proof of Work consensus algorithms.

\section{Related Work}
Consensus shares many properties as convergence to Nash equilibria in various repeated games. Abraham et al. \cite{abraham2011distributed} introduced games that share similarities with the standard consensus problem. We use this work as a motivation for these ideas.

In addition, there has been ample research into scalability and security-first protocol designs. Sompolinsky and Zohar showcase the benefit of this approach in \cite{sompolinsky2016bitcoin, sompolinsky2016spectre, sompolinskyphantom} for different tree and DAG based topologies.

\section{Background}
\subsection{Consensus Games}
We will start with an example of a simple coordination game that motivates the work of the paper, in that it should incentivize the same response (consensus). We will extend this game to its infinite form and analyze the existence of subgame perfect nash equilibria with discounted rewards.
\begin{example}
Suppose that there are $n > 1$ players with action space $A_i=\{0,1\}$.
\begin{itemize}
    \item If everyone plays 0, then $\forall i\in \mathcal{P},~u_i(a) = 1$ where $a=(0,0,\dots,0)$.
    \item If exactly two players $i,j\in \mathcal{P}$ play 1, then $u_i(a)=u_j(a)=2,~u_k(a)=0$ for $k\neq i,j$.
    \item Otherwise, $\forall i\in \mathcal{P},~u_i(a)=0$.
\end{itemize}
\end{example}

From the game above, it is clear that the zero-vector is a pure strategy Nash equilibrium. No player has a unilateral deviation to choose 1, since all players including the deviating party will end up with a payoff of 0. However, two-players who collude and form a deviating coalition have an incentive to choose 1. To that end, we introduce the concept of \textit{resilient} Nash equilibria.
\begin{definition}
Let $G$ be a game. A strategy profile $\sigma=(\sigma_1,\dots,\sigma_n)$ is a \textit{$k$-resilient Nash equilibrium} of $G$, if for all coalitions $C,~|C|\leq k$, and all members $i\in C$, the following condition is satisfied:
\begin{align*}
    u_i(\sigma_C,\sigma_{-C}) \geq u_i(\sigma_C', \sigma_{-C})
\end{align*}
\textit{i.e. coalitions of size at most $k$ do not benefit from deviating} \cite{nisan2007algorithmic}.
\end{definition}
In general, if players are rewarded $\alpha$ for the pure strategy Nash equilibrium of the zero-vector strategy profile and exactly 2 players are rewarded with $\beta$ for playing otherwise as above with $\alpha < \beta$, there will not exist \textit{k-resilient} Nash equilibria for values of $k$ satisfying $\frac{2\beta}{k} > \alpha$. From this simple example, we immediately see the ease and possibility of disrupting consensus that colluding opens up in economically-based, consensus games.

Now suppose we play this game infinitely many times. We would hope to devise a strategy that, in the presence of a colluding coalition of size $k$ satisfying $\frac{2\beta}{k}>\alpha$, is a $k$-resilient subgame perfect Nash equilibrium. Given that the one-shot game is not $k$-resilient, can we devise a way to punish this coalition such that consensus at 0 is incentive aligned? A common, elementary setup in game theory is to find discount factors $\delta$ that incentivize a particular strategy. Although this is unrealistic since players in the real world have arbitrary $\delta$, this information provides insight into the dynamics of the game. The strategy that every honest (non-colluding) player will play is as follows:
\begin{itemize}
    \item Play 0 if every other player played 0 in the previous round.
    \item Play 1 if a colluding party of size $k$ deviates to 1 in the previous round for $t$ rounds.
\end{itemize}
The cooperation strategy is thus a dominant strategy, subgame perfect nash equilibrium strategy for the discount factor $\delta$ satisfying:
\begin{align*}
    \alpha + \delta*\alpha + \delta^2*\alpha +\dots &\geq \frac{2\beta}{k} + \delta^{t+1}*\alpha + \delta^{t+2}*\alpha + \dots \\
    \frac{\alpha}{1-\delta} &\geq \frac{2\beta}{k} + \frac{\delta^{t+1}\alpha}{1-\delta} \\
    \frac{\alpha}{1-\delta} - \frac{\delta^{t+1}\alpha}{1-\delta} &\geq \frac{2\beta}{k} \\
    \frac{\alpha}{1-\delta}\big(1-\delta^{t+1}\big) &\geq \frac{2\beta}{k} \\
    \frac{1-\delta^{t+1}}{1-\delta} &\geq \frac{2\beta}{\alpha k}
\end{align*}
\subsection{Blockchain Terminology}
The important concepts to understand revolve around cryptocurrencies, \textit{Proof of Work} (PoW), and \textit{Proof of Stake} (PoS) protocols. Decentralized cryptocurrencies most importantly solve the double-spending problem. This is the problem in which a single user manages to spend the same cryptocurrency in multiple transactions, a la double spending cryptocurrencies. Since this shouldn't happen in any secure financial system, it remains one of the more important attacks to prevent against.

A PoW system is one in which participants solve some computationally-hard but tractable puzzle. For our purposes, we are interested in the hashing proof of work used by cryptocurrency networks. Traditionally in these proof of work systems, miners deploy massive amounts of computer systems to compete for rewards. Their computational power provides an accurate distribution over their expected rewards and their rewards are paid out without an intentional delay for incentive-compatible reasons (Bitcoin does impose a delay but only for preventing stale/invalid block rewards).

A PoS system is one in which participants stake the underlying blockchain's cryptocurrency to compete for rewards. Rewards should be paid out proportionally to a miner's stake. Within these protocols, new attack vectors are introduced such as the "Nothing at Stake" problem. Since participants use money instead of computer power, they can stake their cryptocurrency on every fork/chain they want, i.e. have nothing at stake for any given chain. To combat this behavior, we impose \textit{slashing conditions} on allowed behavior within a PoS system. When these slashing conditions are satsified, miners lose portions or the entirety of their stake and so are disincentivized to act a certain way.

\section{Model}
Using these concepts as foundations, the synopsis of our model is as follows. A delayed blockchain protocol is one in which rewards are deterministically or randomly delayed to provide a staking mechanism on future payouts. We focus primarily on the notion of double-spending and its potential inclusion within an active blockchain or DAG protocol. The delayed property enables this to be overlaid onto any blockchain network, providing increased security through verifiable \textit{fraud proofs}. Broadcasting valid \textit{fraud proofs} will result in a miner's future rewards being slashed and eliminated, potentially shared with the reporter.
\begin{definition}
A \textit{fraud proof} is an unforgeable, cryptographic proof of misbehavior, submitted to the protocol that can be verified in polynomial time. \\ \\
\textit{Given that double-spends are verifiably erroneous actions (existence of conflicting transactions signed by the same private key), we can transform this into a punishable action. Miners should not attempt to double-spend nor serve users attempting to; any indication signals malice.}
\end{definition}

We adopt the original model from Nakamoto in his seminal paper on Bitcoin and that used by Sompolinsky and Zohar in their work on scalable cryptocurrency protocols. There is a network of miners, represented by a graph $G=(V,E)$ such that each miner has some proportion $p_v$ of the aggregate power, $\sum_{v\in V}p_v = 1$. We will model block inter-arrival times exponentially and model global block arrivals according to a Poisson process $\{N(t)~|~t\geq 0\}$ with rate $\lambda$. A miner $v$'s block creation rate is then $p_v * \lambda$. For simplicity, we sum the transaction fee and block reward into a general reward $\alpha^{(t)}$ for mining new blocks at time $t$. In addition, we denote $G_v^{(t)}=(V^{(t)}, E^{(t)})$ as the network view of miner $v$ at time $t$.
\begin{definition}
Given a cryptocurrency network protocol with reward $\alpha^{(t)}$ and a graph $G$, a miner $v$ can expect to have, at time $t$ with $p=(p_1,\dots,p_{|V|})$ and with cost function $c_{v}^{(t)}(p)$, the \textit{utility} function:
\begin{align*}
    u_v^{(t)}(p) &= (\alpha^{(t)} * p_v * \lambda - c_{v}^{(t)})dt
\end{align*}
\end{definition}

\begin{example}
We begin with a discretized formulation of a delayed mining game. At each discretized time length $\Delta$ starting from 0, a miner is selected and rewarded for mining a block. This miner $v$ has a $p_v$ chance of mining a block each time step. Using the equations defined previously, we can quantify this miner's expected payout at some future time by simply taking a sum over the elapsed time steps.

Upon mining a block, the reward is kept frozen for $k$ time steps. This means that if a miner $v$ mines a block at time $t$, he will receive that reward at time $t+\Delta * k$. The motivation for this, as we briefly mentioned, is to ensure that miner's have stake in the protocol (in their future rewards). In the interval $[t, t+\Delta*k)$, the miner should be properly incentivized not to behave maliciously if he hopes to receive the reward. We can additionally force, as part of the protocol, two additional conditions on participation and rewards to ensure honest behavior is incentive compatible. While these may not be intuitive yet, they will make sense later on.
\end{example}
\begin{definition}
A \textit{$(k,d,\gamma)$-delayed} blockchain or DAG protocol with block and transaction fee rewards $\alpha^{(\cdot)}$ is a protocol where rewards are timelocked for $k$ rounds, $d$ aggregate proof of work is required up front separate from the underlying mining process, and $\alpha^{(\cdot)}$ decays with exponential rate $\gamma$. \\ \\
\textit{Note: $(k,0,0)$-delayed protocols mimic existing protocols --- $k=100$ for Bitcoin and $k=0$ for others.}
\end{definition}
The first two parameters are simple and easily understood as the length of delay in future rewards (which can capture arbitrarily many future rewards) and the startup cost to combat Sybil attacks, respectively. The third parameter, $\gamma$, incentivies miners to "stick around" and actively mine or face arbitrary rates of decay in their future, discounted rewards.
\begin{proposition}
In a \textit{$(k,d,\gamma)$-delayed} protocol with block and transaction fee rewards $\alpha^{(\cdot)}$, timestep lengths $\Delta$, and discount factor $\delta$, it is expected that the utility a miner $v$ has at each round from time $t$ is:
\begin{align*}
    \mathbb{E}_{k,\gamma}[u_v^{((t,t+\Delta))}] = \delta^{k}\Delta(\alpha^{(t)}*e^{-\gamma * \Delta * k} * p_v*\lambda - c_v^{(t)})
\end{align*}
\textit{Note: We will want to find values of $\delta$ such that our enhanced protocol benefits from greater security guarantees as indicated by the potential loss in payouts that come from malicious behavior.}
\end{proposition}

\section{Delayed Blockchains}
Recall that the goal of these changes is to provide increased security and incentive compatibility to arbitrary protocol designs. We want to prevent against double-spend attacks from large mining parties up to a majority. Currently, the state of the art in \textit{Proof of Work} protocol design has no punishment capabilities by design. Rewards are paid out more or less instantly; delays are not used to incentivize honest behavior but as a check on stale blocks.

In blockchain networks, we also have access to a robust public/private key infrastructure. Actions by mining participants in the network are uniquely signed by participants' private keys. If we hope to punish malicious participants, we would hope to "blacklist"/punish their public identities (public keys) within the protocol. In the current iteration of public blockchains however, this suffers from the following drawbacks.
\subsection{Drawbacks of existing protocols}
\begin{enumerate}
    \item Creating new public/private key pairs or identities is trivial. Malicious miners can continuously change to a newly generated identity.
    \item Payouts are instantaneous. Assuming a particular chain is not reverted, large mining parties can withdraw funds immediately.
    \item Large parties do not have proportional "skin in the game" or stake against their mining power.
\end{enumerate}
For the purpose of this paper, we address the three issues above. The usefulness of a $(k,d,\gamma)$\textit{-delayed} blockchain protocol allows us the ability to tune the severity of the punishment to arbitrary strengths in order to combat the mitigation techniques above. The intuition for each is as follows:
\begin{enumerate}
    \item If $d>>0$, creating new public/private key pairs to behave maliciously with old key pairs should be disincentivized. In addition, if $\gamma$ grows exponentially as a function of the number of times a past winning miner drops out of the current miner set, past winners will lose substantial portions of their future rewards. In many cases, $\gamma$ will incentivize sticking around for sufficiently large $k$ such that $d$ can be set to 0.
    \item Payouts are no longer instantaneous and miners cannot instantly withdraw funds until they honestly cooperate with the protocol rules for a parameterized amount of time.
    \item Miners have stake in reporting malicious behavior and in behaving according to the protocol.
\end{enumerate}

\section{Subgame Perfect Equilibria}
As we showed in our toy example, there are lengths of time for which deviations are not rational decisions. We operate strictly in a rational agent model and leave notions of \textit{Byzantine} or irrational behavior for future work. Towards similar results, we consider a double-spend deviation on a $(k,d,\gamma)$\textit{-delayed} blockchain with deviation payoff $\epsilon>0$. That is, an adversary can double-spend and receive an arbitrarily large reward of $\epsilon$. We would expect that a double-spend attempt should only be a rational deviation if $\epsilon$ is greater than the aggregate payout of a miner's future rewards.

\subsection{$(k,0,\gamma)$\textit{-delayed}} For simplicity we assume there is no cost to mining or Sybil startup cost. We also assume that all miners have been playing for sufficiently long. The expected payoff for a miner $v$ with power $p_v$ and discount factor $\delta$, who has honestly participated in the protocol for $l>>k$ time steps is:
\begin{align*}
    u_v^{(1:l)}(p) = \sum_{i=1}^l \delta^{k+i-1}(\alpha*e^{-\gamma * k} * p_v*\lambda)
\end{align*}
If the payoff for double-spending is $\epsilon$. Then the payoff for double spending after $l$ rounds in the infinitely repeated game can be represented by replacing the last $k$ expected payoffs with $\epsilon$. This is because we assume another miner submits a fraud proof and slashes the $k$ most recent, expected payouts:
\begin{align*}
    u_v^{(1:l-k)}(p) + \epsilon + u_v^{(l:\infty)}(p)
\end{align*}
Then it follows, a profitable attack would satisfy the following:
\begin{align*}
    u_v^{(1:\infty)}(p) &\leq u_v^{(1:l-k)}(p) + \epsilon + u_v^{(l+1:\infty)}(p) \\ 
    u_v^{(l-k+1:l)}(p) &\leq \epsilon \\
    \sum_{i=l-k+1}^l \delta^{k+i-1}(\alpha*e^{-\gamma * k} * p_v*\lambda) &\leq \epsilon \\
    \epsilon &= O(k*\alpha*\delta^k*e^{-\gamma * k}*\lambda)
\end{align*}
We arrive at the big-O form using a rudimentary simplification of the left-hand sum. We know that $\delta\in (0,1)$, therefore, for $l>>k,~\delta^l < \delta^k$. Each term of the left-hand sum has a $\delta^{l'},~l'\geq l$ and so is bounded by $\delta^k$. The sum is taken over $k$ terms, so we add a factor of $k$ in front. Additionally, since $p_v\in [0,1]$, it follows that its omission can only add to the growth of the right-hand term.

If $\epsilon=O(k*\delta^k*e^{-\gamma * k})$, then a double-spend attack should be a profitable deviation; strict equality would indicate a guaranteed profit. Given we set $d=0$, there is no cost to getting punished and malicious miners can trivially jumpstart mining on new public/private key pairs. Now, if we increase $d>0$, the cost of the double spend will increase. This, however, has no effect on the payoff of honest participation. Therefore, for $\epsilon$ sufficiently large, the protocol designer can tune $d$ to push attack costs higher.

To analyze subgame perfect equilbria, we must reverse the inequality. We leave the equations unsimplified and closed forms for future work. These bounds serve to illustrate the various ways we can increase attack costs without altering honest rewards.

\subsection{$(k,d,\gamma)$\textit{-delayed}} Now consider the case where $d>>0$. For a given miner $v\in V$ with power fraction $p_v$, we let $r_v^d$ denote the random variable of the total number of rounds (resp. time interval length) of the blockchain protocol that it takes miner $v$ to solve $d$ proof of work. It follows that $\mathbb{E}_{v,d}[r^d_v]<\infty$, since proofs of work for hashing live in a finite space.

Now, when an attacker $v$ double-spends, their address is invalidated or "blacklisted" by the protocol's rules. In order to startup again, the attacker $v$ generates a clean public/private key pair and shifts $p_v$ to this new pair. The miner $a$ must complete $d>>0$ proof of work before earning any rewards. Then it is only profitable to do so if the following is satisfied:
\begin{align*}
    u_v^{(1:\infty)}(p) &\leq u_v^{(1:l-k)} + \epsilon + u_v^{(l+r_v^d + 1:\infty)}(p) \\
    u_v^{(l-k+1:l+r_v^d)} &\leq \epsilon \\
    \sum_{i=l-k+1}^{l+r_v^d} \delta^{k+i-1}(\alpha*e^{-\gamma * k} * p_v*\lambda) &\leq \epsilon \\
    \epsilon &= O((k+r_v^d)*\alpha*\delta^k*e^{-\gamma * k}*\lambda)
\end{align*}
With this, we only increase the cost of a successful double-spend attack. Since honest, long-lived miners are rewarded with only needing to solve the startup proof of work once, they benefit in the long run. The right balance between tuning these parameters would be an assignment that reaches a balance between the value of future discounted rewards and the cost of successfully executing a double-spend attack.

\section{$\gamma$ selection}
Using $\gamma$ as intended would require protocol upgrades. If at any time $t$, we have a view of all miners or stakers by a list of public/private key identities $pks^{(t)}=\bigoplus_{v\in V^{(t)}} pk_v$, then as long as the winner $v^*$ at time $t$ continues to actively mine for $\Delta * k$ more time steps, they can expect their future reward to decay more slowly or not at all compared to the event that they stop any time before the delayed interval has passed.

More generally, $\gamma$ could be parameterized to the frequency of participation of a particular miner. Newly generated addresses will experience fast decay in their future rewards compared to incumbent, "reputable" addresses. Short-lived miners can be penalized for their undesirable behavior, while long-lived nodes can be incentivized through proper $\gamma$ selection. In any of these distributed systems, we want to incentivize long-lived participation and truthful participation. Imposing restrictions on the behavior the protocol designer desires allows the designer control of the underlying protocol's incentives.

\section{Implementation}
Without detailing exactly how these ideas can be implemented, we offer a variety of high-level techniques. For a $(k,d,\gamma)$\textit{-delay} implementation, we can extend the mining protocol to include timelocked transactions for winners of the mining game, parameterized by a length $k$. The underlying protocol can maintain a set $pks^{(t)}$ at each time $t$ and track a variety of information about all miners it has seen in the past (miners can reach consensus on this dataset as well) for $\gamma$ funtionality. Additionally, a double proof of work layer can be added to the traditional mechanics to include functionality of the startup cost, $d$, parameter.
\section{Conclusion}
In this paper, we presented a new technique for improving the security of any Proof of Work blockchain/DAG protocol. The main technique involves delaying the rewards of miners to create a Proof of Stake style staking mechanism. We impose slashing conditions around \textit{fraud proofs} that can be determined by protocol designers. By slashing future rewards, we simulate slashing a miner's stake in the system.
\bibliographystyle{unsrt}
\bibliography{bibliography}
\end{document}